\documentclass[aps,floatfix,twocolumn,footinbib,superscriptaddress,showpacs]{revtex4}
\usepackage{epsfig}
\usepackage{amsmath,amssymb}
\usepackage{bm}
\begin{document}

\title{Electric control of ferromagnetism
in Mn-doped semiconductor heterostructures}


\author{Christian Ertler\footnote{email:christian.ertler@uni-graz.at}}

\author{Walter P\"otz}
\affiliation{Institute of Theoretical Physics, Karl-Franzens University Graz,
Universit\"atsplatz 5, 8010 Graz, Austria}


\begin{abstract}

The interplay of tunneling transport and
carrier--mediated ferromagnetism in narrow semiconductor
multi--quantum well structures containing layers of GaMnAs is
investigated within a self-consistent Green's function approach, accounting for disorder in the Mn--doped regions 
and unwanted spin--flips at heterointerfaces on
phenomenological ground.  
We find  that the magnetization in GaMnAs  layers can be
controlled  by an external electric bias.  The underlying mechanism is  identified as spin--selective hole  tunneling in and out of the
Mn-doped quantum wells,  whereby the applied bias determines both hole population and spin polarization in these layers. 
In particular we predict that, near resonance, ferromagnetic order in the Mn doped quantum wells is destroyed. 
The interplay of both magnetic and transport
properties combined with structural design potentially leads to
several interrelated physical phenomena, such as  dynamic spin
filtering, electrical control of magnetization in individual
magnetic layers, and,  under specific bias conditions, to
self--sustained current and magnetization oscillations (magnetic
multi-stability).   Relevance to recent experimental results is discussed.

\end{abstract}

\maketitle

\section{Introduction}

Electric control of magnetism in nanostructures must be viewed as an
important milestone  on our road map for successful realization of
spintronic devices. Although most of the operations in such devices
ultimately should be based on spin--only processes, i.e., processes
not associated with (highly dissipative) electric charge transport,
to gain best benefits from such designs, spin must be manipulated
both during the input, control, and read--out stage and eventually be
coupled to charge.  Several schemes achieving this goal have been
explored both at the quantum and semi-classical level, such as the
electric distortion of the orbital wave function of spin carriers in
inhomogeneous (effective) magnetic fields \cite{Shin2010:PRL},
electric g-tensor control \cite{Roloff2010:NJP,Kroutvar2004:N}, or
spin torque transfer.\cite{Myers1999:S, Ralph2008:JMMM,Wenin2010:JAP} 
Here we explore, on theoretical grounds,  the influence of an electric bias on the ferromagnetic state 
and  feasibility of electric control of ferromagnetism in 
Ga$_{1-x-y}$Al$_y$Mn$_x$As multiple quantum wells.  Structural design, including effective potential profiling and 
doping to position emitter and collector quasi--Fermi levels, as well as 
tunneling is used to control hole density and spin polarization within the Mn doped layers.

Dilute magnetic semiconductors (DMS) have been realized by doping of
conventional ZnS--structured  semiconductors with elements providing open
electronic $d$ or $f$ shells.  This  has added yet another degree of
freedom to the rich spectrum of physical phenomena in semiconductors
available for material design with potential for technological applications.\cite{Jungwirth2006:RMP} 
A prominent example is bulk Ga$_{1-x}$Mn$_x$As where Mn on the Ga sites
provides both an open d-shell with a local magnetic moment and
a hole which may establish ferromagnetic ordering among the
Mn d--electrons, a mechanism known as carrier--mediated
ferromagnetism.\cite{Ohno1996:APL,VanEsch1997:PRB,Dietl2000:Science}
The preferentially anti--parallel alignment  of the 3/2  spin of the
mobile holes with the 5/2 spin of the localized Mn d--electrons
promotes ferromagnetic ordering of the latter below a critical
temperature of  up to  $\sim 150$ K.  
Theoretical work  has confirmed strong hybridization between the 5d Mn and  3p electrons 
in the ground state.\cite{Jain2001:PRB}
The effective
hole--concentration--dependent exchange field lifts the spin
degeneracy of the holes' energy bands and thus goes hand in hand with hole spin polarization. 
The Mn ions sitting on Ga sites act as acceptors and are believed to give rise to acceptor levels which lie about 100 meV above the valence band edge.\cite{Jungwirth2006:RMP,VanEsch1997:PRB,Schneider1987:PRL}
Photoluminescence experiments indicate the co-existence of 
holes bound to Mn sites and itinerant holes which participate in establishing magnetic order amongst the Mn ions below $T_c$.\cite{Sapega2009:PRB}

Since structural defects  of bulk and confined layers of Ga$_{1-x}$Mn$_x$As  depend on grow conditions, Mn concentration $x$ , and annealing procedures it is not too surprising that experiments have come up with somewhat different  conclusions regarding the ``electronic structure of bulk Ga$_{1-x}$Mn$_x$As".  More recent work seems to hint at the existence of an impurity band which forms at Mn concentrations above ~ 1.5~\% leading to a metal--insulator transition in high--quality GaMnAs.\cite{Burch2006:PRL,Richardella2010:S,Ohya2011:NP} 
The Fermi level in these samples is reported to lie in the impurity band 
and  the valence--band properties remain largely GaAs--like.\cite{Ohya2011:NP} 
The radius of the Mn acceptor wave function has been  measured to be about 2~nm, indicating that Mn$_{Ga}$ is not a shallow acceptor.\cite{Richardella2010:S}
In contrast, other studies rather hint at a disordered top valence 
band edge containing the Fermi energy, but no isolated impurity band is present.\cite{Jungwirth2006:RMP}   Recent theoretical work has led to the conclusion  that  a tight--binding approach (within the coherent-potential approximation for disorder) and local--density functional theory + Hubbard U correction cannot account for an isolated impurity band.\cite{Masek2010:PRL} 
Other theoretical work has lead to the conclusion that  disorder may enhance ferromagnetic stability.\cite{Lee2007:PRB,Berciu2002:PhysicaB}
Ionized impurity scattering seems to play the dominant role in explaining Hall resistivity data.\cite{Yoon2004:JAP}

Controlled growth of heterostructures containing  crystalline  layers of GaMnAs of high structural quality has remained a challenge up to date.  Nevertheless, tunneling spectroscopy has confirmed size quantization effects in GaMnAs quantum well layers.\cite{Ohya2007:PRB,Ohya2010:PRL,Ohya2011:NP}   However, compared to crystalline GaAs well layers, in an otherwise identical structure,  the signature appears to be rather weak and in no sample yet, apparently, has one observed negative differential conductivity due to resonances associated with GaMnAs well layers.   This hints at a significant concentration of defects, reminiscent of thin layers of amorphous Si where similar transport studies have revealed size quantization effects but, to our knowledge, not negative differential conductivity.\cite{Miyazaki1987:PRL,Li1993:PRB}    
 Experimental evidence indicating a coexistence of localized and extended Bloch--like states in bulk GaMnAs, in general, allows
 the prediction that, in thin layers, certainly for $\leq 3$ nm,  of GaMnAs extended states will be subjected to confinement effects 
(quantization and energy shifts) while localized states will remain largely unaffected.  This is similar to external magnetic--field effects on point defects or quantization effects in amorphous Si.\cite{Poetz1983:SSC, Li1993:PRB}
Assuming that no significant additional defects arise  in GaMnAs heterostructures,  this makes plausible experimental reports on quantum confinement effects arising from (ferromagnetic) GaMnAs layers in thin heterostructures.\cite{Ohya2007:PRB,Ohya2010:PRL,Ohya2011:NP}
Indeed, when one succeeds to incorporate high--quality magnetic layers
in semiconductor heterostructures  strongly
spin-dependent carrier transmission can be predicted due to spin-selective
tunneling.\cite{Sankowski2007:PRB} In magnetic resonant tunneling
structures of high structural quality this spin splitting may be used for a realization of spin
valves, spin filtering, and spin switching devices
\cite{Likovich2009:PRB,Slobodskyy2003:PRL,Slobodskyy2007:APL,Petukhov2002:PRL,Ohya2007:PRB,Ohya2010:APL,Ertler2006a:APL,Ertler2007a:PRB}, all representing  important
ingredients for spintronic-based device technology.

In several experiments ferromagnetism has been generated in bulk GaMnAs 
by,  electrically or
optically, tailoring the hole density.\cite{Ohno2000:N, Boukari2002:PRL} 
In 2d-confined systems
containing layers of Ga$_{1-x}$Mn$_x$As  the magnetic order depends
strongly on the local spin density, which can be influenced by the
tunneling current, resulting in a bias-dependent exchange
splitting.\cite{Dietl1997:PRB,Jungwirth1999:PRB} 
A spin-density dependent
exchange splitting in ferromagnetic structures enriches the dynamic
complexity by offering a mechanism for  external electrical
control of the ferromagnetic state.
This is in contrast
to structures comprising paramagnetic DMS, such as ZnMnSe, in which
a giant Zeeman splitting of the bands is induced by applying an
external magnetic field of the order of a few Tesla. 

Already nonmagnetic multi--well heterostructures exhibit interesting
dynamic nonlinear effects which are based, however, on different physical
mechanisms, such as the formation of electric field domains and the
motion of charge dipoles through the structure.\cite{Eaves1989:SSE, Poetz1990:PRB, Stegemann2007:NJP,Bonilla2005:RPP} 
Recently it has
been predicted that, in heterostructures containing paramagnetic DMS
wells, this kind of phenomena can be controlled by an external
magnetic field.\cite{Sanchez2001:PRB, Bonilla2007:APL,Escobedo2009:PRB}  Using an incoherent, sequential tunneling model
we have proposed that {\em ferromagnetic} multi--well structures can
generate ac spin currents, a phenomenon which originates from
time--dependent inversion of the spin population in adjacent
wells.\cite{Ertler2010:APL}

In this article we investigate spin--selective hole transport in
GaAs/AlGaAs/GaMnAs heterostructures within the limit of moderately thin samples
with predominantly {\em coherent} transport characteristics.  We apply a
non-equilibrium Green's function formalism based on a tight--binding
Hamiltonian for the electronic structure, including self-consistency
regarding the charge density and the exchange splitting of the
effective potential, as well as charge transfer to the contacts.
Both the carriers' Coulomb interaction and the exchange coupling
with the magnetic ions are described within a mean-field picture.
Details of our model are exposed in Sect.~\ref{sec:model}.  The
mechanism of electric control of magnetization switching is
explored for two generic structures containing, respectively,  one
and two layers of Ga$_{1-x}$Mn$_x$As.  Results are given in
Sect.~\ref{sec:results}. We also provide a qualitative explanation
for the occurrence of spin-polarized current oscillations, predicted
in an earlier paper \cite{Ertler2010:APL}, and investigate the
influence of spin flip processes at the interfaces on the total
current spin polarization.  Since disorder seems to play a major role in actual samples 
we study the effect of substitutional  disorder on a qualitative level and discuss the robustness of the effects predicted here.  
Relevance to experiment is discussed.   In particular, we can give an explanation for the absence of exchange splitting (magnetization) under resonance bias condition reported in a recent experiment and identify 
characteristic features which may be explored in future experiments.  
Summary and conclusions are given in Sect.~\ref{sec:sum}.

\section{Physical Model}\label{sec:model}

The magnetic semiconductor heterostructure  is described by a two--band tight-binding Hamiltonian
for the heavy holes $(J_3=\pm 3/2)$.  It is
given in the form
\begin{eqnarray}
H_s &=& \sum_{i,\sigma} \varepsilon_{i,\sigma} |i,\sigma\rangle\langle
i,\sigma|\nonumber\\
&&+\sum_{i,\sigma\sigma'}t_{i,\sigma\sigma'}|i,\sigma\rangle\langle
i+1,\sigma'|+ \mathrm{h.c.},
\end{eqnarray}
where  $\varepsilon_{i,\sigma}$ is the spin-dependent ($\sigma
=\uparrow,\downarrow \equiv \pm 1$) onsite energy at lattice site $i$, 
$t_{i,\sigma\sigma'}$ denotes the hopping-matrix between
neighboring lattice sites, and $\mathrm{h.c.}$ abbreviates the Hermitian conjugate term. Spin conserving hopping gives a diagonal
matrix $t_{i,\sigma,\sigma'} = t\delta_{\sigma\sigma'}$, whereas spin flip processes can be taken into account
by introducing off-diagonal elements. The hopping parameter $t = -\hbar^2/(2 m^* a^2)$ depends on the effective mass $m^*$ and the lattice spacing $a$ between to neighboring
lattice sites. The onsite energy
\begin{equation}
\varepsilon_{i,\sigma} = U_i - e \phi-\frac{\sigma}{2}\Delta_i
\end{equation}
includes the intrinsic hole band profile $U_i$ due to the band
offset between different materials, the electrostatic potential
$\phi$ with $e$ denoting the elementary charge, and the local
exchange splitting $\Delta_i$.  Near the band-edges this model is
equivalent to an effective--mass model, however, it has the
advantage that structural imperfections and spin--flip processes can
readily be incorporated.  Moreover, it can be extended to arbitrary
sophistication by introducing a larger set of basis
functions.\cite{Sankowski2007:PRB, Schulman1983:PRB,Poetz1989:SM,DiCarlo1994:PRB}

Within a mean-field approach the exchange coupling between holes and magnetic
impurities can be described by two interrelated effective
magnetic fields,
respectively, originating from  a nonvanishing mean spin
polarization of the ions' d--electrons  $\langle S_z\rangle$ and from the hole
spin density $\langle s_z\rangle = (n_\uparrow-n_ \downarrow)/2$.\cite{Dietl1997:PRB,Jungwirth1999:PRB, Fabian2007:APS}  The exchange splitting of the hole bands is then given
by
\begin{equation}\label{eq:delta}
 \Delta(z) = -J_\mathrm{pd} n_\mathrm{imp}(z) \langle S_z\rangle(z)~,
\end{equation}
with $z$ being the longitudinal (growth) direction of the structure,
$J_\mathrm{pd} > 0 $  is the coupling strength between the impurity
spin and the carrier spin density (in case of GaMnAs p-like holes
couple to the d-like impurity electrons), and $n_\mathrm{imp}(z)$ is
the impurity density profile of magnetically active ions. Since the
magnetic order between the impurities is mediated by the holes, the
effective impurity spin polarization depends on the mean hole spin
polarization via
\begin{equation}\label{eq:Szgen}
  \langle S_z\rangle= - S B_S\left( \frac{S J_\mathrm{pd} \langle s_z \rangle}{k_B T}\right),
\end{equation}
where $k_B$ denotes Boltzmann's constant, $T$ is the lattice
temperature, and $B_S$ is the Brillouin function of order $S$, here with
$S = 5/2 $ for the Mn impurity spin. Combining Eq.~(\ref{eq:delta}) and
Eq.~(\ref{eq:Szgen}) leads to a self-consistent effective Hamiltonian
for the holes $H_\mathrm{eff} = -\sigma \Delta(z)/2$ with
\begin{equation}\label{eq:delta1}
 \Delta(z) = J_\mathrm{pd} n_\mathrm{imp}(z)
S B_S\left\{\frac{S J_\mathrm{pd} [n_\uparrow(z)-
n_\downarrow(z)]}{2 k_B T}\right\}.
\end{equation}
Note that in thermodynamic equilibrium of a quasi 2D-systems, such
as a quantum well,  the hole spin density polarization $\langle
s_z\rangle$ is the key figure of merit for the appearance of
ferromagnetism.

Within a Hartree mean-field picture space-charge effects are taken into account
self-consistently  by calculating the electric potential from the Poisson equation,
\begin{equation}\label{eq:poisson}
 \frac{\mathrm{d}}{\mathrm{d}z} \epsilon \frac{\mathrm{d}}{\mathrm{d}z}\phi =
e\left[ N_a(z) - n(z)\right],
\end{equation}
where $\epsilon$ denotes the dielectric constant and $N_a$ is the
acceptor density. The
local hole density at site $|i\rangle$ is given by
\begin{equation}\label{eq:n}
 n(i) = \frac{-i}{A a}\sum_{k_{||},\sigma}\int\frac{\mathrm{d}E}{2\pi} G^<(E;i\sigma,i\sigma)~,
\end{equation}
with $A$ being the in-plane cross sectional area of the structure, and $k_{||}$
denotes the in-plane momentum. The non-equilibrium ``lesser''
Green's function $G^<$ is calculated from the equation of motion
\begin{equation}\label{eq:gless}
 G^< = G^R\Sigma^<G^A
\end{equation}
where $G^R$ and $G^A = [G^R]^+$ denotes the retarded and advanced Green's function, respectively.
The scattering function $\Sigma^<=\Sigma^<_l+\Sigma^<_r$ describes the inflow of
particles from the left $(l)$ and right $(r)$ reservoirs \cite{Datta:1995}
\begin{equation}
\Sigma^<_{l,r} = f_0(E-\mu_{l,r})(\Sigma^A_{l,r}-\Sigma^R_{l,r})~,
\end{equation}
where $f_0(x) = [1+\exp(x/k_B T)]^{-1}$ is the Fermi distribution function and
$\mu_l$ and $\mu_r$, respectively, denote the quasi--Fermi energies in the contacts.
The retarded and advanced self-energy terms $\Sigma^R = \Sigma_l^R+\Sigma_r^R$ and $\Sigma^A = [\Sigma^R]^+$
account for the coupling of the system region to the left and right semi--infinite chains, for which an
analytic expression can be derived.\cite{Datta:1995, Economou:1983}
The retarded Green's function is then given by
\begin{equation}\label{eq:gr}
 G^R = \left[E+i\eta-H_s-\Sigma^R\right]^{-1}~,
\end{equation}
with $i\eta$ being a positive infinitesimal imaginary part of the
energy.

Together with adjusting the Fermi energies relative to the band
edges in the leads to ensure asymptotic charge
neutrality \cite{Poetz1989:JAP}, the band splitting given by
Eq.~(\ref{eq:delta1}),
 the Poisson equation Eq.~(\ref{eq:poisson},\ref{eq:n}),
and the kinetic equations Eqs.~(\ref{eq:gless}) and (\ref{eq:gr})
have to be solved self-consistently until convergence to a
steady--state solution is reached. Nonlinearities in both Hartree
and exchange term can give rise to multi--stable behavior, as will
be discussed below.  If this selfconsistency loop terminates with ferromagnetic ordering in the system, the effective one--particle potential  
is different for spin--up and spin--down holes, thus leading to spin filtering in transmission. 

After obtaining the self-consistent potential profile
the spin-dependent transmission probability $T_{\sigma'\sigma}(E)$
from the left to the right reservoir, as a matrix element of the structure's S-matrix,  can be calculated
from special matrix elements of the retarded Green's function \cite{DiCarlo1994:PRB}
\begin{equation}
T_{\sigma'\sigma}= T_{\sigma'\leftarrow\sigma}(E) =
\frac{v_{r,\sigma'}|G^R(E;r\sigma',l\sigma)|^2}{v_{l,\sigma}
|G^0(E;l\sigma,l\sigma)|^2}
\end{equation}
with $G_0$ denoting the free Green's function of the asymptotic
region, and $v_{l,\sigma}$ and $v_{r,\sigma}$, respectively, are the
spin-dependent group velocities in the leads.
$G^R(E;r\sigma',l\sigma)$ is computed most conveniently by adding
one layer after another which requires merely 2x2 matrix inversions
for the present two--band model.\cite{Economou:1983}

Finally, the steady--state current
is obtained from scattering theory (generalized Tsu-Esaki formula),
\begin{eqnarray}
 j_{\sigma'\sigma} & = & \frac{e m^* k_B T}{(2\pi)^2\hbar^3}
\int_0^\infty \mathrm{d} E\: T_{\sigma'\sigma } g(E)\nonumber\\
g(E) & = & \ln\left\{\frac{ 1 + \exp\left[(\mu_l-E)/k_B T\right]}{
1 + \exp\left[(\mu_r-E)/k_B T\right]}\right\}.
\end{eqnarray}
The applied bias $V=(\mu_l-\mu_r)/e$ is determined by the difference in quasi-Fermi levels of the contacts.

We would like to point out that we conduct a genuine non--equilibrium study whereby the quasi--Fermi level positions are associated with the contacts.  Self--consistency then leads to an effective, in general, spin--dependent one--particle potential.  Thus one is not confronted with the question where to place the Fermi level in the GaMnAs layers.   Essential to confinement effects  is the existence of states near the top of the valence band edge of GaMnAs which have a coherence length of at least the layer thickness.   Highly localized states,  whether separated from or attached to the top valence band edge, will not be very sensitive  to 
finite layer width.  
While in the bulk and thermal equilibrium the itinerant hole exchange model firmly relates hole density to T$_c$ and the Fermi energy, in a non-equilibrium tunneling situation this is different.   The key question is whether tunneling can induce a net hole spin polarization or not.  As is shown below, we find that this depends on structural properties as well as on the applied bias.


\section{\label{sec:results} Results}

We start with a symmetric double--barrier structure containing a  single GaMnAs  quantum well
and investigate the role of resonant hole tunneling on the magnetic state of the device.
For the simulation we use generic parameters for GaMnAs and GaAs:
 $m^* = 0.4\:m_0$, $\epsilon_r = 12.9$, $V_\mathrm{bar} = 400$ meV, $\mu_l =\mu_r= 80$ meV, $d = 20 \AA$,
$w = 25 \AA$, $n_\mathrm{imp} = 1\times10^{20} $cm$^{-3}$,
$J_{\mathrm{pd}} = 0.15$ eV nm $^3$ \cite{Lee2000:PRB}, $T = 4.2$~K,
where $m_0$ denotes the free electron mass,  $\epsilon_r$ is the
relative permittivity, $V_\mathrm{bar}$ is the bare barrier height
of AlGaAs relative to GaAs, $d$ and $w$, respectively, are the
barrier and quantum well width. 
The thermal equilibrium position of the Fermi energies $\mu_l =\mu_r$ was deliberately chosen close to the first resonance to promote ferromagnetic 
ordering in the well region at zero bias. 
The background charge $N_a$ is
assumed to be only about 10\% of the Mn doping $n_\mathrm{imp}$
since GaMnAs is a heavily compensated system, most likely  due to Mn interstitial
or antisite defects.\cite{VanEsch1997:PRB,DasSarma2003:PRB}
The hole densities in the quantum well can be adjusted by the Mn doping level and the quasi-Fermi levels in the contacts. 
As can be seen from Eq.~(\ref{eq:delta1}) the exchange splitting increases with the hole density in the case of a steady
particle spin polarization. The value of the exchange coupling constant varies in literature to some extend
$J_{\mathrm{pd}}\approx 0.04 - 0.16$ eV nm. Since we use an optimistic value for $J_{\mathrm{pd}}$, we assume only moderate Mn$_\mathrm{Ga}$ doping in the well. Higher Mn$_\mathrm{Ga}$ densities and smaller values of $J_\mathrm{pd}$ will give  very similar results.  
 
Disorder effects in the GaMnAs layers are modeled by performing a configurational average over structures with randomly
selected onsite and hopping matrix elements of the tight-binding
Hamiltonian in the Mn doped region. For each specific Hamiltonian the transport problem (I-V curve) is solved self--consistently.  The 
final result is obtained by averaging over all configurations.   Typically 300 configurations are used for one I-V curve.
For the numerical simulation we assume a fixed 5\% Mn concentration in the well and model substitutional disorder.  If a Mn ion is present at a given  lattice site
in the well the onsite energy is shifted according to a Gaussian distribution 
around  a mean onsite energy--shift of 40 meV and a standard deviation of 20 meV, which are reasonable
values according to experimental results which indicate either an impurity band slightly above the valence band edge or a defect--induced  valence band tail.\cite{Ohya2011:NP,Richardella2010:S,Masek2010:PRL}  
The nearest--neighbor hopping matrix elements for such a site are sampled according to a Gaussian between  5\% and 25\% standard deviation
($\sigma_t$) of its bulk value $t$. This model for substitutional disorder leads to a hybridization of quantum confined hole states, associated with bulk like valence band states, and localized defect states arising from Mn$_{Ga}$ sites.   The degree of hybridization depends on 
layer thickness since it controls the position of the quantized heavy hole band relative to the energy of the localized Mn acceptor levels.    This hybridization and the experimentally found Mn acceptor radius of about 2 nm calls for rather thin GaMnAs layers to ensure quantization effects in transport.\cite{Ohya2011:NP}  
 
\begin{figure}[!t]
\centering
\includegraphics[width=0.95\linewidth]{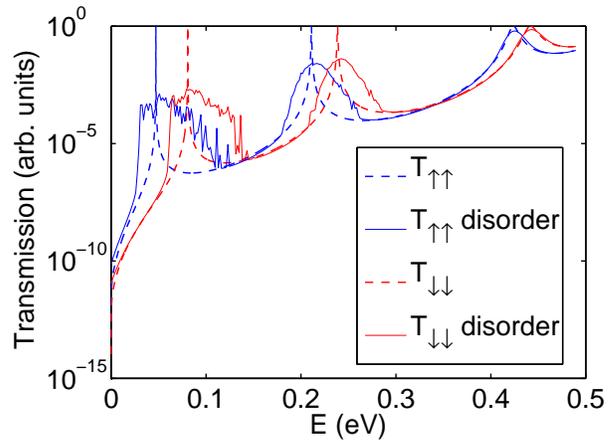}
\caption{(Color online) Spin-dependent transmission probability of the double barrier structure at zero bias with
and without disorder ($\sigma_t = 5 \%$).}
\label{fig:T}
\end{figure}


The calculated spin-filtering effect via distinct tunneling probabilities for
spin-up and spin-down holes arising  from the exchange term is
displayed in Fig.~\ref{fig:T} in which  the transmission
probability at zero bias is plotted versus energy of incidence $E$.    This figure also gives a qualitative account of the density of states in the GaMnAs well region discussed above.  
For an idealized GaMnAs layer which,  at the valence band edge, is modeled as a GaAs layer plus exchange term,  one obtains sharp spin doublets 
which are exchange--split  by  about 25-30 meV (see dashed versus solid lines in Fig.~\ref{fig:T}).  
The state of zero spin polarization of holes represents  an unstable
equilibrium since, below T$_c$, the slightest perturbation in
spin--polarization drives the system into a partially ordered lower
energy state (spontaneous symmetry breaking) due to the exchange
interaction.  The latter, in turn,  accounts for different
effective barrier profiles for spin--up and spin--down holes.  It is
this nonlinear effect that can be utilized to control the hole spin
polarization and thus the favorable Mn spin orientation by
structural design and applied bias.  Placing the Fermi level near the first heavy--hole resonance 
promotes this effect, similar to  the formation of Cooper pairs near the Fermi edge of an interacting electron gas.

Spin--selective tunneling into and out of the Mn doped wells, regardless of whether sequential or resonant, promotes hole spin polarization and, thus, alignment of the Mn spins as long as spin-depolarizing processes in the heterostructure are slow compared to the effective tunneling rates.  Furthermore, disorder which leads to spectral broadening of the resonances may suppress spin--selective tunneling.   Inspection of Fig.~\ref{fig:T} shows an asymmetric broadening and significant overlap of the 
transmission peaks under substitutional disorder, modeled as discussed above, which is particularly pronounced for the first heavy--hole resonance since it is most sensitive to potential fluctuations.  The asymmetric (``anti--bonding") shift towards higher energies is due to  hybridization with 
Mn acceptor levels below the conduction band edge.   The latter do not contribute to resonant transport.  Even at the moderate disorder for the effective hopping matrix element of 5 percent, a significant 
overlap in spin-up and spin--down resonance is obtained.  Increased disorder and/or spin--flip scattering will eventually wash out spin--selectivity in transmission and a destruction of ferromagnetic ordering under bias  must be expected  since {\em unpolarized} holes are steadily fed into the GaMnAs regions.   Exchange splitting at zero bias for 5\% and 25\%, respectively,  is reduced  to 33~meV and 23~meV.

Clearly, our effective one--dimensional modeling of (substitutional)  disorder must be viewed as a limited estimate since it corresponds to a  cross--sectional average of  transport though 
uncorrelated effective linear chains.  Correlations from disorder parallel to the heterointerface will play a role in the establishing of coherence and ferromagnetic order in real structures relative to the idealized 
homogeneous mean--field model adopted here, since both ferromagnetic order and disorder effects are highly dependent upon spatial dimensionality.\cite{Kaxiras:2003, Ashcroft:1976}  Additional types of disorder from Mn clustering, Mn interstitials, etc. may be present in real structures.  The role of disorder in the formation of ferromagnetic order in diluted magnetic semiconductors has been explored theoretically and, remarkably, certain form of disorder has been predicted to promote ferromagnetic ordering.\cite{Lee2007:PRB, Berciu2002:PhysicaB}  
In experiment, STM studies have given information on the nature of defects near the surface of GaMnAs samples.\cite{Burch2006:PRL,Richardella2010:S}

\begin{figure}[!t]
\centering
\includegraphics[width = 0.9\linewidth]{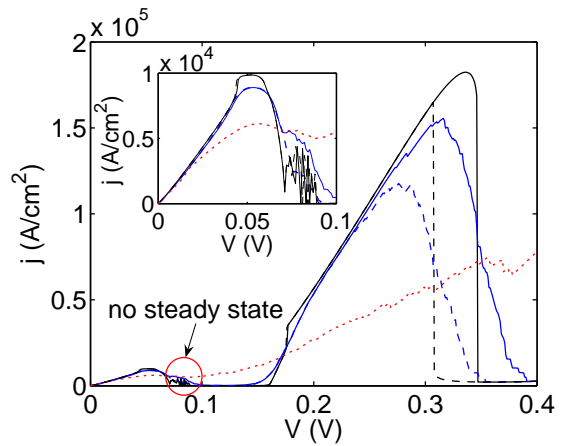}
\caption{(Color online) $IV$-characteristics of a magnetic double barrier structure due to heavy--hole associated bands with and
without disorder. The solid (dashed) lines refer to a voltage up (down)-sweep without disorder (black lines) and moderate disorder of $\sigma_t = 5 \%$ (blue lines). The $IV$-curve is flattened at higher defect concentrations as indicated by the red dotted
line  $\sigma_t = 25 \%$ assuming a voltage up-sweep.
As shown in the inset in the voltage range of $V = 0.07 - 0.09$ V for the case of a high-quality sample no steady state is
reached, suggesting the occurrence of dynamic effects.}
\label{fig:IVgreen}
\end{figure}

\begin{figure}[!t]
\centering
\includegraphics[width= 0.9\linewidth]{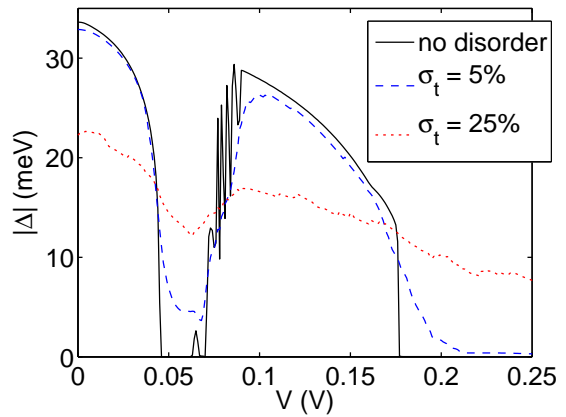}
\caption{(Color online) The configuration averaged spin splitting $|\Delta|$ in the quantum well
 as a function of the applied bias with and without disorder.}
\label{fig:Delta}
\end{figure}

\begin{figure}[!t]
\centering
\includegraphics[width=0.9\linewidth]{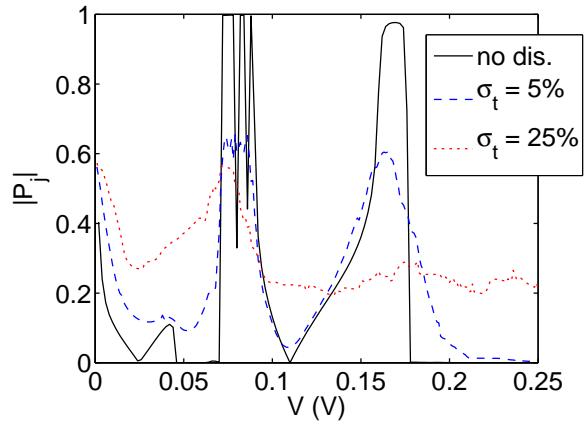}
\caption{(Color online) Averaged current spin polarization $|P_j|$ versus
applied bias $V$ with and without disorder.}
\label{fig:Pjdis}
\end{figure}

The current-voltage I--V characteristics, plotted in
Fig.~\ref{fig:IVgreen}, reveals the typical hysteretic behavior of
resonant tunneling diodes for an up- and down-sweep
of the applied bias. This well known intrinsic bistability
occurs due to different charging of the well depending on the
bias--sweep direction.   Since our model ignores contributions from the light--hole band, associated resonances 
are missing in the plot.  The latter are important due to in--plane non--parabolicity effects in narrow layers,  however, low--lying resonances associated with heavy and  light holes generally are  clearly 
separated in energy.  For the present structure a light--hole--band resonance would be expected between the first two heavy--hole--associated resonances, thus strongly reducing the peak--to valley ratio and  contributing spin $\pm 1/2$ holes to the Mn--doped layers.   It should be observed that only single resonance peaks are observed in the I--V characteristics in spite of spin doubles in the (zero--bias) transmission spectra.  Furthermore, the drop in 
current beyond the first heavy--hole peak value (see insert in Fig.~\ref{fig:IVgreen}), unlike in ballistic models for nonmagnetic tunneling structures (see second peak), is gradual even in the absence of disorder.  This broadening of the resonance can be attributed to ferromagnetic ordering away from the first resonance peak which tends to widen the bias window for meeting the resonance condition.

Disorder effects  diminish the peak--to valley ratio  but
regions of negative differential resistance are maintained for weak  disorder.   Since experiments have not shown negative differential conductivity in such a structure we have increasing disorder and find its disappearance at relatively high
hopping disorder of about $\sigma_t = 25$~\% (see Fig.~\ref{fig:IVgreen}).   This indicates that defects other than Mn acceptors are present in real samples.
Further numerical studies regarding this issue will be published elsewhere.\cite{Ertler2011:JCE}

In Fig.~\ref{fig:Delta} the  average
exchange band spin splitting $|\Delta|$ in the quantum well, characterizing its magnetic state,  is plotted versus applied bias.  It shows that ferromagnetism can be controlled by the applied bias in this
structure  near the first current peak, remarkably, even when disorder is sufficiently strong to suppress negative differential conductivity. 
At zero bias ferromagnetic ordering is energetically preferred since the Fermi $\mu_l=\mu_r$ level is located close to the edge of the first heavy--hole subband.  As the bias is increased tunneling into the upper doublet state 
becomes allowed from the emitter side reducing the net hole--spin polarization (in spite of increasing hole density) and the effective exchange field decreases to zero and both spin--up and spin--down subbands go into resonance.  Note that under moderate bias both emitter and collector contribute to the population of the well region.  As the bias increases resonant population from the emitter gets shut off and hole polarization is determined by the collector leading once more to a build-up of the exchange field for a bias regime between 0.08~V and 0.2~V, when finally the collector quasi-Fermi level drops below the hole subbands and the well region becomes almost depleted of holes.  For higher bias no further 
spontaneous magnetization has been obtained within our self--consistency loop.  The overall feature of the bias dependence of the 
exchange splitting thus somewhat resembles its behavior versus temperature, with ``$T=T_c$" corresponding to a bias of about 0.18~V.   It arises from the fact that it is the number of {\it spin--polarized} holes which determines the maximum spontaneous magnetization for given Mn$_Ga$ concentration.  A simple model for the dependence of the Curie temperature in resonant tunneling systems has been given by one of us before.\cite{Ertler2008:APL} The voltage-dependence of the Curie temperature under resonant tunneling has also been studied before.\cite{Ganguly2005:PRB}

The displayed build--up and
destruction of ferromagnetic order as a function of applied bias
can be further understood by the exchange interaction which is mediated by
spin--polarized holes.  In an ideal 2D particle system with parabolic dispersion there is no energy gain by magnetic ordering due to the constant density of states associated with each spin subband: energy gained by lowering one subband is exactly cancelled by raising the other.    
However, here we deal with  a 3D heterostructure which
favors a spin ordered state when the quasi--Fermi level lies 
near (within about half of the maximal exchange splitting)  the bottom of a well subband resonance. 
If the temperature in the
contacts is sufficiently low, one subband after the other will go
through resonance.  Thus, when only the lower spin--subband is in
resonance holes in the magnetic well will tend to be be spin--polarized.
However, as bias is increased eventually the subband with opposite
spin orientation will also go into resonance thus reducing spin
polarization and magnetic ordering in the GaMnAs layer.    
When, for a given bias,  the well region cannot be populated  (lack of hole density of states) or no energy gain can be drawn from ferromagnetic ordering,   loss of the latter
will result.

\begin{figure}[!t]
\centering
\includegraphics[width=0.9\linewidth]{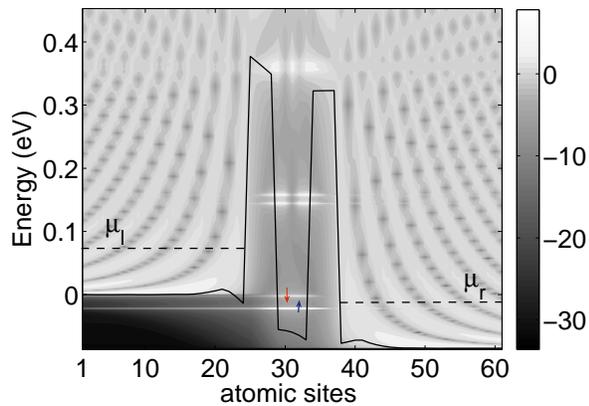}
\caption{(Color online) Logarithmic local density of states (LDOS)
as a function of energy at the bias $V = 0.085$ V. The
self-consistent band profile is indicated by the solid line. The
spin-splitting of the quasi--bound states is clearly visible.}
\label{fig:ldos}
\end{figure}

Interestingly, in the voltage range of $V = 0.07 - 0.09$ V no steady
state solution can be found for the low disorder sample case. Instead the solution for the
magnetization is oscillating, as shown in Fig.~\ref{fig:Delta}, suggesting the occurrence of dynamic
effects. This behavior can be understood qualitatively as follows:
Figure~\ref{fig:ldos} shows a contour plot of the local density of
states, for an applied bias
$V=0.085$ V lying in the critical voltage range. The self--consistent band profile is indicated by the solid line.
For the emitter Fermi energy of
$\mu_l = 0.08$ eV only the two
ground state (potentially spin--split) subbands in the quantum well
participate in the tunneling transport.  At this bias condition and
hole spin polarization the lowest (spin--up) subband may be
populated by holes from the collector side, whereas the spin--down
level is almost empty since it cannot be reached elastically by
either emitter or collector. Since the (steady--state) band
splitting $\Delta$ is proportional to the spin polarization
$(n_\uparrow - n_\downarrow)$ the well magnetization increases with
spin polarization, pushing the spin--down level upwards in energy.
At some point holes can start to tunnel from the emitter side into
the spin--down level.  This in turn decreases the total spin
polarization  and, hence, effectively pushes the spin--down level
back  below the emitter's band edge. From there, the process starts
anew, leading to an oscillatory behavior in well magnetization,
tunneling current, and spin polarization.\cite{Ertler2010:APL}

\begin{figure}[!t]
\centering
\includegraphics[width=0.9\linewidth]{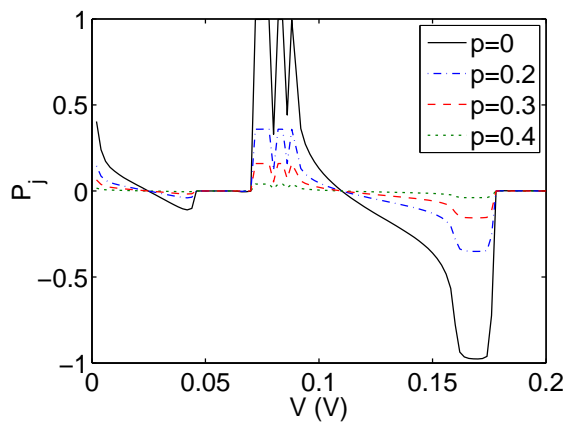}
\caption{(Color online) Current spin polarization $P_j$ versus
applied bias $V$ taking into account spin flips at the
hetero-interfaces. The polarization is diminished for an increasing
spin flip probability $p$ becoming unpolarized for $p = 1/2$.}
\label{fig:pflip}
\end{figure}

Although the I-V curves in Fig.~\ref{fig:IVgreen} do not display spin-split resonance peaks,  but merely a broadening of the resonance, 
the steady--state current at low bias is spin polarized as shown in Fig.~\ref{fig:Pjdis}.
As the bias is increased from zero, current spin polarization is reduced and reversed before it drops to zero through resonance.  Above resonance current spin polarization reemerges (due to the action of the collector)
and once more changes sign before dropping and remaining at zero in one-to-one agreement with the behavior of the exchange field.   Although resonance peaks in the I-V curve my be suppressed by disorder, see Fig.~\ref{fig:IVgreen}, 
the bias--dependence of spin polarization in the current may persist and may be observed in experiment as a bias--dependent spin valve.

In order to study qualitatively the influence of spin flip processes at the
hetero--interfaces on the total current spin polarization at the collector side, $P_j =
(j_{\uparrow\uparrow}+j_{\uparrow\downarrow}-j_{\downarrow\uparrow}-j_{\downarrow\downarrow})/j$
with $j = \sum_{\sigma\sigma'} j_{\sigma\sigma'}$, we introduce
off-diagonal hopping matrices $V_{i,\sigma\sigma'}$ in the
tight-binding Hamiltonian. In general for $N$ interfaces there are
$2^N$ different flip configurations.  For each of them a simulation
is performed and the results are finally averaged by weighting with
the probability for the occurrence of the configuration. In the case
of a double-barrier structure we have four hetero-interfaces, giving
$16$ configurations. However, flipping at the first barrier
interfaces is inefficient, since it does not change the total
current or spin polarization. Single flipping at the third or fourth
interface does also not modify the total current density but inverts
the spin polarization to $-P_j$. By introducing single spin flip
probabilities $p_i, (i=1,\ldots,N)$ at the interface $i$, the
probability of a flipping process at the second barrier is then
given by $p_{\mathrm{flip}} = p_3(1-p_4)+(1-p_3)p_4$. Hence, the
mean spin polarization results in
\begin{equation}
\langle P_j\rangle = P_j(1-2p_\mathrm{flip}).
\end{equation}
The bias-dependent current spin polarization for different spin flip
probabilities (assuming $p_3 = p_4 = p$) is plotted in
Fig.~\ref{fig:pflip}. The spin polarization decreases  for
increasing $p$ with $\langle P_j\rangle [p] = \langle
P_j\rangle[1-p]$ reaching its minimum $\langle P_j\rangle = 0$ for $p
= 1/2$.  From this analysis we conclude that our results will not be altered significantly when the spin--orbit interaction is taken into account in the analysis.  

\begin{figure}[!t]
\centering
\includegraphics[width=0.95\linewidth]{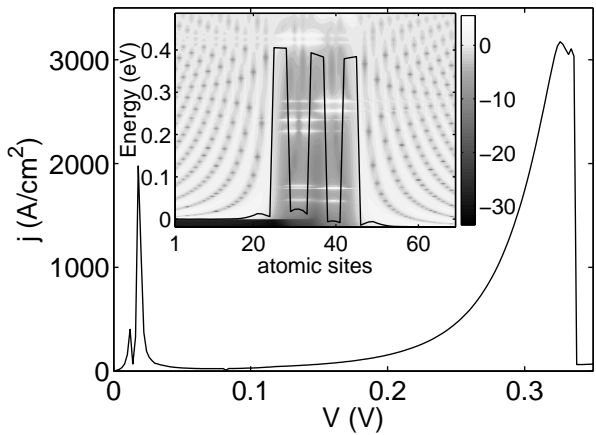}
\caption{IV-characteristics of a three-barrier
structure with two coupled quantum wells made of GaMnAs. At the
current maxima resonance conditions are fulfilled, i.e., the
quasi--bound states of the adjacent wells become energetically aligned.
The inset shows the local density of states at the applied bias $V =
0.03$ corresponding to the first current maxima. } \label{fig:IVqws}
\end{figure}

\begin{figure}[!t]
\centering
\includegraphics[width=0.9\linewidth]{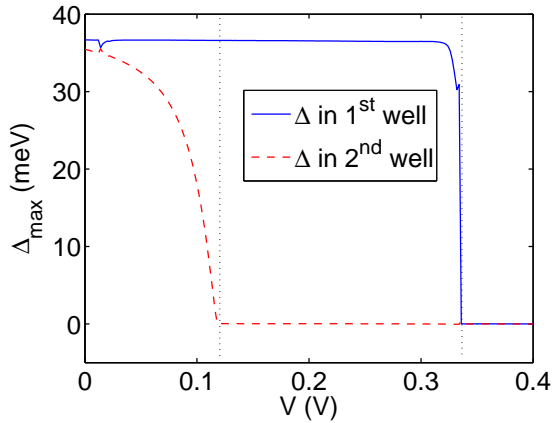}
\caption{(Color online) Maximum exchange splitting
$\Delta_{\mathrm{max}}$ in the first (solid) and second well (dashed
line) as a function of the applied bias.} \label{fig:Dqws}
\end{figure}

While spin--selective hole tunneling may allow electric control of ferromagnetic order, tunneling spectroscopy, in turn, provides a sensitive experimental tool  for exploring the electronic structure of 
mesoscopic semiconductor systems.\cite{Smoliner1996:SST}
Recently, tunneling spectroscopy experiments have been performed on thin layers of GaMnAs.\cite{Ohya2011:NP}   The authors have verified ferromagnetic ordering in their samples 
 (with Mn concentration of typically $x\approx $5 to 15~$\%$ and layer thickness ranging from 4 to 20 nm)
and have measured their respective Curie temperature.  Their measurements indicate that Mn induced defect states remain separated from the GaAs--like valence band edge as evidenced by a pinning of the Fermi level.   Furthermore, they find clear signatures of quantization effects in the transmission spectra of their samples and report an absence of spin-splitting in the resonances which they can fit to a GaAs-like k.p model, including light-hole states.  We believe that these experimental findings compare favorably with the general features of our results.   Moreover, we have provided an explanation for the observed absence of  ferromagnetic ordering near resonance in spite of ferromagnetic behavior of the sample at zero bias.
It would be interesting to perform spin--sensitive tunneling spectroscopy on these samples since, according to Fig.~\ref{fig:pflip},  such a measurement gives more detailed information about the bias dependence of ferromagnetic ordering than the I-V curve and its derivatives.   This can test the prediction that ferromagnetic order which can be achieved at zero bias is destroyed near resonance would be verified, and that electric switching back and forth between the ferromagnetic and paramagnetic state can be achieved.  

We now explore feasibility of selective magnetization switching
among several magnetic layers of high structural quality.  We investigate a three-barrier
structure with two adjacent GaMnAs quantum wells, choosing an
asymmetric structure with the second well being thinner than the
first one ($w_1 = 25 \AA$, $w_2 = 20\AA$).  All other parameters are
as in the previous structure. Quantum confinement gives rise to a
higher ground state energy in the second quantum well at zero bias.
The resonant alignment of the ground state subbands of the two wells
is therefore achieved at a finite voltage as shown in the inset of
Fig.~\ref{fig:IVqws}, corresponding to the first maximum  in the
current-voltage characteristics at about $V=0.02$ V which is plotted
in Fig.~\ref{fig:IVqws}. The second current maximum result from the
resonance of the first excited state subbands of both wells. Next to
possible exchange splitting the finite separating barriers cause the
energy levels in the two quantum wells to further split into bonding
and antibonding subband states. However, for our structure the
middle barrier is too thick and the natural energy broadening of the
quasi--bound states is too large for resolving this additional
splitting in the local density of states.

Having two coupled quantum wells allows one to realize several
magnetic configurations. The maximum (steady--state) exchange
splitting of the two wells as a function of the applied bias is
plotted in Fig.~\ref{fig:Dqws}, revealing three different regions.
For low voltages both wells are magnetized due to the build--up of
spin polarization in the wells due to resonance of the ground state
subband levels with populated reservoir states.  Exchange causes a
relative shift in the  density of states for spin--up and spin--down
holes which, in turn, stabilizes ferromagnetism in both layers. When
the second well goes off resonance at around  $V > 0.03$ V the
accumulated spin polarization in the second well is preserved, since
for voltages up to $V \approx 0.1$ V the collector Fermi energy
$\mu_r$ is still higher than the bottom of its ground state subbands
thus maintaining spin polarization. For voltages in the interval
$0.12$ V$ < V < 0.33$ V the first well remains magnetic, whereas the
second well becomes nonmagnetic, since the ground state subbands are
no longer filled from the collector side.
 At sufficiently high
bias,   $V > 0.33$ V,  also the first well becomes demagnetized
since holes can no longer resonantly populate its two lowest (now
degenerate) subbands from either emitter or collector thus resulting
in a completely nonmagnetic structure.

Several simplifying assumptions have been made in the present
analysis which should, just as well as experimental aspects,  be
discussed. The present model is based on an effective-mass-like
two--band approach for the heavy holes in the structure.   This
approximation should at least  qualitatively be correct since the
applied bias is kept below typically  0.2~V and most of the phenomena
discussed here occur at lower bias.   Thus it can be expected, that
this model describes  effects qualitatively correct.    We are
currently working on more realistic tight--binding formulations
using a significantly increased number of basis states in
conjunction with density functional plus dynamic mean-field models
to arrive at a more detailed and realistic electronic
structure.\cite{Chioncel2011:PRB}  
 Impurity scattering effects  have been accounted for on a phenomenological level within the TB model.
Our ballistic model neglects
electron--phonon scattering within the heterostructure altogether
and the electron--electron interaction is described within
mean--field theory.  In thin structures, such as the ones studied
here where effective tunneling rates are higher than carrier--phonon
scattering rates and optical phonon transitions are suppressed
energetically the former assumption should be rather well fulfilled
and not alter significantly subband population within the
heterostructure.   Electron--electron scattering may play role,
however, as long as it does not involve spin--flip processes should
not influence our basic conclusions much. 

 Clearly the effects studied here require low temperatures,
for one to favor ferromagnetic ordering and, secondly, to preserve
strong hole--spin polarization in the carrier injection process.  It
is well known that, at least at low temperatures, structural
imperfections are the main source for reduction of nonlinear
effects, such as the peak--to--valley--ratio in the IV
curve.\cite{Poetz1989:SM, Poetz1989:SSE, Chevoir1990:SS,
Mizuta:1995} It is most likely the difficulty in clean sample
preparation which has slowed experimental progress on thin--layer
semimagnetic semiconductor heterostructures.   High quality doping
profiles and high quality interfaces must be achieved within one
growth process.\cite{Ohya2007:PRB,Likovich2009:PRB,Ohya2010:APL}
Growth of good quality DMS layers needs low temperature molecular
beam epitaxy which, however,  adversely  affects interface quality.
Usually thin GaAs spacer layers are inserted to smooth the
surfaces.\cite{Ohya2007:PRB,Ohya2010:APL}  Furthermore, GaMnAs
layers must be thick enough to support ferromagnetism.
Qualitatively, all structural imperfections lead to broadening of
resonances.  Once the latter becomes comparable to the (theoretical)
maximum  of the exchange energy induced spin--splitting,
spin--selective tunneling and, hence, tunneling--induced control of
magnetic ordering may be suppressed.   Even in the presence
of disorder, as long as it does not go hand in hand with strong
spin--flip processes, achieving bias--control of hole--spin polarization in
the GaMnAs layers should allow one to manipulate magnetization.

\section{Conclusions and Outlook}\label{sec:sum}

In summary, we have used a ballistic steady--state transport model
to investigate bias--induced magnetic multi--stability in
AlGaAs/MnGaAs quantum well structures. Ferromagnetic exchange, as
well as the hole Coulomb interaction are treated within
self--consistent mean--field approximation.  Substitutional disorder is treated phenomenologically within a tight--binding model.

Our studies indicate  that in these systems ferromagnetic ordering
can be controlled selectively by an externally applied bias.  The
underlying mechanism is found in spin--selective tunneling due to
the anti--ferromagnetic exchange interaction between itinerant heavy
holes and localized Mn d--electrons. In structurally  suitably
designed heterostructures the applied electric bias allows control
of the ferromagnetic state, as well as electric and  spin current
density.    

 In the simplest structural case in form of a double barrier structure containing a GaMnAs well, we predict that ferromagnetic ordering in the well, when present at zero bias, is lost under bias near the first  heavy-hole resonance, allowing a switching back and forth between the magnetic and nonmagnetic state in the well.    In GaMnAs multi-well structures we predict that the loss of ferromagnetic order can be engineered structurally to occur at different applied bias for the individual layers.

Within our model we are able to provide a possible explanation for the absence of exchange splitting near resonances, 
as observed in recent tunneling spectroscopy measurements on thin GaMnAs layers.\cite{Ohya2011:NP}   We generally predict that ferromagnetic order which may be achieved in 
GaMnAs quantum well layers under zero bias tends to be destroyed under resonance condition since the well region then is swept by unpolarized holes.   Under favorable conditions detailed in the main text, 
ferromagnetic order may be reestablished above resonance.  Such a behavior should be revealed experimentally by  spin--sensitive tunneling spectroscopy.\cite{Ando2005:APL}
  
In previous work based on a complementary time--dependent sequential tunneling
model including intra--well scattering we have predicted that,  under specific bias conditions,  the interplay of transport and
magnetic properties can result in robust self-sustained charge and
magnetization oscillations.\cite{Ertler2010:APL}   The present model, albeit based on the resonant--tunneling picture,  
backs the possibility of such phenomena by predicting bias regions in which no steady--state solution for the current exists.  

Disorder and spin--flip effects have been modeled on a phenomenological level.  We find that disorder due to Mn taking a Ga site alone should not  suffice to destroy spin--selective tunneling, nor should spin flips at a rate expected in these structures, for example from the spin--orbit interaction, significantly suppress spin polarization of the steady--state current.  As expected, our analysis does show that disorder and spin flip processes do reduce  the total average current spin polarization, however, not as efficiently as the resonance peaks in the I--V curve.

We conclude that multi--well structures containing GaMnAs layers
may allow one to realize various bias-dependent magnetic configurations.
While the current investigation of bias induced effects considers
only bias in longitudinal direction, i.e., a 2--terminal
configuration, applying additional gates in transverse direction
(multi--terminal configuration)  should allow for an additional
control knob to move spin--split subbands in and out of resonance with the contact states and/or to inject spin--polarized holes into the Mn--doped regions.  Such a structure has been studied in a
recent experiment.\cite{Ohya2010:APL}

\section{Acknowledgment}

This work has been supported by the FWF project P21289-N16.


\end{document}